# MAINWAVE: Multi Agents and Issues Negotiation for Web using Alliance Virtual Engine

Debajyoti Mukhopadhyay, Saurabh Deochake, Shashank Kanth,
Subhadip Chakraborty, Suresh Sarode

Maharashtra Institute of Technology, Pune 411038
India

{debajyoti.mukhopadhyay, saurabh.deochake, kanthshashank, subhadip.schakraborty,
sarodesuresh.r}@gmail.com

**Abstract:** This paper showcases an improved architecture for a complete negotiation system that permits multi party multi issue negotiation. The concepts of multithreading and concurrency has been utilized to perform parallel execution. The negotiation history has been implemented that stores all the records of the messages exchanged for every successful and rejected negotiation process and implements the concepts of artificial intelligence in determination of proper weights for a valid negotiation mechanism. The issues are arranged in a hierarchical pattern so as to simplify the representation and priorities are assigned to each issue, which amounts to its relative importance. There is refinement of utilities by consideration of the non-functional attributes. So as to avoid overloading of the system, a maximum number of parties are allowed to participate in the entire mechanism and if more parties arrive, they're put into a waiting queue in accordance to certain criteria such as the first come first serve or the relative priorities. This helps in fault tolerance. It also supports the formation of alliances among the various parties while carrying out a negotiation.

**Keywords:** Concurrent Programming, Distributed/Internet based software engineering tools and techniques, Electronic commerce, Network repositories/data mining/backup, Artificial Intelligence Introduction

## 1 Introduction

Negotiation is a process in which different agents attain an agreement on a joint future venture. The gradual rate of increase in the number of transactions executed by means of electronic channels such as the internet in the B2B and the B2C E-commerce has credited to the need of improved negotiation models and formally semantic protocols. Negotiation is a crucial necessity for B2B e-commerce activity. The current E-commerce activities are no longer limited to the business organizations but it takes into account various methodologies such as advertisements of products endorsing their features and the various categories in the internet, providing a platform for generation and exchange of offers and counter offers for each of the parties i.e. the buyers and the sellers and the alliances that could be ad-hoc or permanent. This whole procedure requires strict cohesion to an appropriate protocol, sets of rules and particular threshold for every negotiation agreement. The pre-existing negotiation models are often incapable in providing the essential support for executing multi party multi issue negotiations.

An agent is an encapsulated computer system based on an unknown environment with the ability of elasticity, self governing action in that environment in order to meet its design objectives.

Negotiation objects possess wide vibrancies of issues which have to be fulfilled for a specific agreement to be established. These objects may consist of only one specific issue (such as price), or, the issues might even be multiple at times, in fact in multitudes (viz. quality, timings, penalties etc.). In order to simplify and systematically align the representation, the attributes are ordered in a hierarchical manner. The model proposed in this paper provides multi-party multi-issue negotiation.

Concurrent activities exist or occur at the same time. Our model presents a concurrent solution to the execution of various negotiations at the same time i.e. multiple agents belonging to different categories negotiate over multiple issues at the same time in parallel. A mathematical analysis has also been suggested for assignment of weights and calculation of a depreciation factor to be discussed later. The negotiation protocol allows prioritization of attributes, taking into consideration a utility function which considers the non functional attributes and assigns weight in accordance to the relative importance of the features.

The negotiation history is a database that stores all the sets of messages that are exchanged among different parties while the performance of any negotiation. Hence it is a store house of every step that is executed in attaining a successful or failed negotiation. The negotiation history helps in the logical implementation of a concept of artificial intelligence known as "learning by experience". This results in the selection of weights for each issues which are the most beneficial for the current deal by analyzing all the previous interactions and hence reaching a experience-wise profitable decision.

Alliances are formed when two or more parties join hands together in pursuit of some common business aims and then make deals as a unified body. MAINWAVE allows a special parameter allies described later in the paper that is used for the formation of alliances during negotiation process.

Plug-ins are external components consisting of mathematical and business logic which can be imported to the agent's interface as per the requirements of processing while performing business and logical calculations for determining the proper weights and generation of offers and counter offers. Plug-ins help in the extensibility, flexibility, modularity and wide spectrum of calculation possibilities which are helpful in the establishment of a logical foundation to the entire negotiation process.

There is always a possibility of the system to be overloaded when the number of parties exceeds a threshold boundary. Hence fault tolerance is observed in our model that helps in the manifestation of a waiting queue. Whenever the participant parties exceed a pre-stated number, the emerging new parties are restricted access from the negotiation market, and are stored in the waiting queue in accordance to criteria's such as the first come first serve or priority based queuing.

Thus, our paper proposes MAINWAVE: Multi Agents and Issues Negotiation for Web using Alliance Virtual Engine. The paper is structured as follows: Section 2 gives an outline of the existing work in this field, Section 3 describes of the model proposed in this paper, Section 4 includes the salient features of our model, Section 5 provides the future scope, and finally, we conclude this paper in section 6.

## 2 Related Work

In order to settle an agreement, the negotiation objects or the issues on which the An overview of the process of negotiation can be represented by a buyer and a seller agent with their respective deadline $t_{max}$ by which the agreement has to be reached, for finalizing the deal. Multiple issues such as price, quality, etc. have to be considered and once the buyer or the seller finalizes the deal, it declines offers from all others. Three approaches have been used: Game theory, Heuristics and Argumentation Approach.

Game theory [2] attempts to capture behavior in strategic behavior situations, where each player's success in making a decision depends on the decision of others. A protocol based on this approach is simple to implement but the biggest problem with this is that it is computationally complex. The number of calculations increases exponentially in each round. Another problem is that it is only suitable for specialized models that are used for specific types of interdependent decision making.

Some of the limitations of game theory based techniques can be overcome by making use of heuristics [1]. Here, the agreement space can be searched in a non-exhaustive fashion. Heuristics are experience based techniques that make educated guesses. So, what we get are good results and not optimal results. The problems with using heuristics are that, (1) the outcomes are sub-optimal as they do not consider the full space of possible outcomes and (2) it is impossible to predict how the system and its constituent agents will behave in different circumstances so it requires extensive evaluation.

Finally, Argumentation Framework [7] applies logic based reasoning and argumentation in the proper design of agents for web services. Thus, we propose a negotiation model that attempts to arrive at a compromise among the above three approaches and is computationally simple at the same time.

To implement an efficient negotiation system, there should be capable subcomponents of the negotiation system. The main types of the negotiation services that could be included are as given in [6]: (1) Fully-delegated negotiation services: "Fully delegated" can be defined as a term that prescribes the incorporation of automated negotiation agents in the system. (2) Interoperable messaging system: This messaging system has the possibility of being the message broker. (3) Negotiation process management: The negotiation process management forces the negotiation rules that should be carried out while in the negotiation. In order to aid an open and dynamic negotiation environment [9], the requirements for a product specification language are: (1) Formal Semantics: There is a requirement of a language with formal semantics to accommodate heterogeneity of participants. (2) Dynamic re-configurability: Dynamic changes are allowed to be incorporated into the syntax through this feature. (3) Compatibility: Protocols can be designed to follow a modular style using this. Compatibility permits the construction of negotiation protocols that incorporate independent negotiation. (4) Composability: using runtime leads to the option of dynamically adding new protocols switching among them. Our model allows fully-delegated negotiation services by using automated agents, negotiation process management and formal semantics by enforcing fixed communication primitives and compatibility by defining a uniform interface which can be implemented using web services.

Interaction among agents requires a uniform messaging system or a standard interface which will allow communication using fixed message primitives. T. D. Nguyen and N. R. Jennings in their paper for concurrent, bilateral negotiation [3] make use of the following message primitives: Offer (a proposal made by one agent to the other), Counter-offer (a revised proposal from an agent in response to a proposal it has received), Accept (accept a proposed offer), Finalize (finalize a deal with the chosen seller and vice versa), Decline (reject the temporarily accepted previous offer), Withdraw (terminate the negotiation thread). The difference between accept and finalize is that a buyer may accept several offers from multiple sellers in any one negotiation episode by making use of heuristics presents two types of agent viz. buyer and seller. We are adopting these primitives for inter-agent communication in our model.

Using game theory or heuristics approach, three strategies commonly adopted as given in [3] are: (1) conceder (2) non-conceder and (3) linear. A fourth strategy presented in implementing privacy negotiation [10] to conclude the negotiation process is also worth exploring.

Additional challenges in implementing a negotiation framework include the discovery problem [8], i.e., discovery of potential agents to negotiate with. To solve this we make use of a centralized negotiation server, which will provide directory services for agent discovery.

The main purpose of every agent in Kasbah e-marketplace [4] is to finalize an acceptable deal which is in accordance with the conditions and the constraints specified by the buyers and the sellers. The Kasbah e-marketplace is amongst the very first initiatives at exploiting agent technology for automated negotiations in e-Commerce. An entire amalgam of buyer and seller agents meets at the centralized Kasbah e-marketplace. These agents logically seek out potential buyers or sellers and negotiations are then carried out among the selected parties on behalf of their owners. Unfortunately, the Kasbah agents can only negotiate over the single issue of price which is the biggest disadvantage.

Overall the following are the shortcomings of the existing negotiation protocols: (1) non-consideration of non-functional attributes (2) restricted to single issues (3) No hierarchical representation of negotiation issues. Here, we have proposed "MAINWAVE," a new model that aims to address these issues in a computationally efficient manner.

## 3   Our Model

In this paper, we propose an enhanced model that is composed of the negotiation system and the participant agents in the entire mechanism. The negotiation system initializes a negotiation process that utilizes the concept of weighted utility wherein the non-functional attributes are considered. The negotiation system consists of three main components: (1) advertisement repository and (2) condition-checker (3) Alliance engine.

The actual negotiation starts as the agents submit their advertisements into the advertisement repository. The advertisement repository is the key database that stores all the agent specific information, such as the product id, the agent ip, the product name, the allies parameter and the threshold counter. The conditional checker then executes the conditional matching of the buyer and the sellers according to the content of the advertisements and classifies them with respect to their unique product id. On satisfying the condition of the presence of at least one buyer and one seller for a particular product, the control then passes to the agents who now carry out the negotiation by establishing a negotiation market. The negotiation process involves the generation of offers and counter-offers for a specified period of time and it terminates either when an agreement has been attained or when the threshold counter expires. The negotiation system is consequentially alerted.

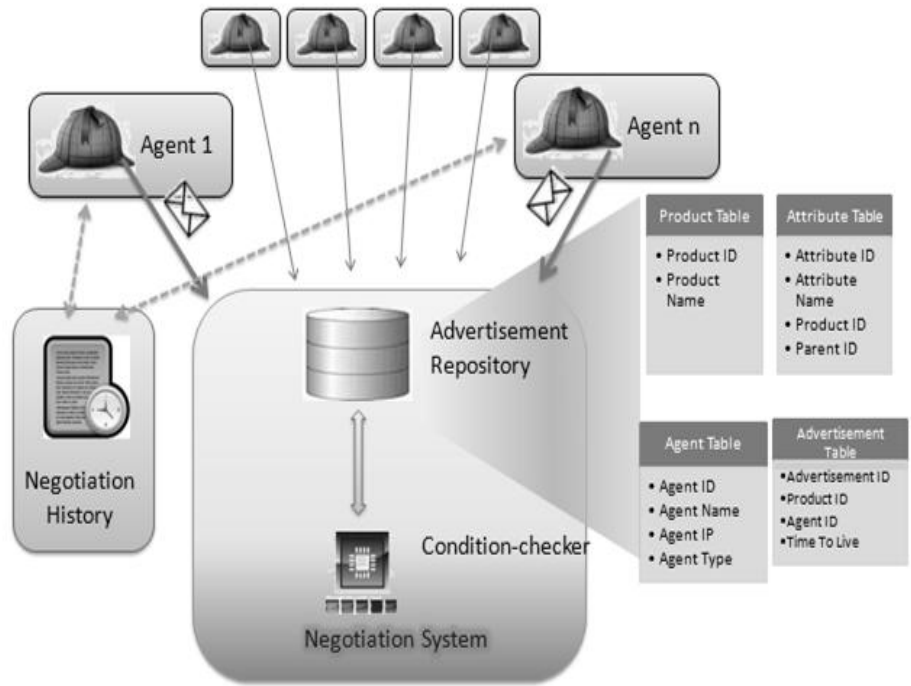

**Figure 1 Model Architecture**

### 3.1 Negotiation System

The negotiation system has the following components: (1) advertisement repository and (2) condition checker (3) negotiation history (4) Alliance Engine.
The advertisement repository which is our main database comprises of the following tables: (1) Agent table: It stores the detailed information of every individual agent viz. the agent id, name, IP, allies and type of agent (buyer or seller). (2) Product table: it contains details of products on which the negotiation is to be performed. It consists of two fields: product id and product name. (3) Attributes table: Each product has various attributes which can be aligned in a hierarchical manner so that it can be classified into proper sub-sections and which will simplify the representation. (4) Advertisement table: It comprises of advertisements in strict adherence to a specific template and formal semantics. All the necessary details that summarize the agent and the products that would be put up for negotiation are included in the advertisement table. The attributes of this table are advertisement id, product id, agent id and validity counter. (5) Ongoing Negotiation table: This is a market for the entire negotiation that is maintained by the agent and includes the agent id and the product id and the number of offers that have been generated. The primary purpose of this table is to ensure that when there is an ongoing negotiation between two or multiple agents on a particular product and a new agent is willing to negotiate on the same product, then that agent is allowed to be added directly to the same negotiation process so as to support dynamic re-configurability.

### 3.2   Condition Checker

The condition-checker scans the advertisements in the advertisement repository and performs matching and classifying the advertisements in accordance to their product ids. The condition checker then searches for the presence of at least one buyer and one seller for the particular product type and if they are present, then it begins the negotiation process. The control then passes entirely to the agents thereafter.

### 3.3   Negotiation History

Our model has the capability of keeping track of all the information and data shared among the participant parties by setting up a database that stores the records of every negotiation sequence carried out in the system. The negotiation history stores all the sets of messages which have been shared between the inter- communicating parties while performing a negotiation. This helps in the concept of "learning by experience" which is an application of artificial intelligence. It helps in assignment of weights to the issues which would result in beneficial achievement of deals at the current stage of the undergoing negotiation. These assignments are done by analyzing the previous sets of successful negotiation cases and determining the pivotal steps that led to the success.

### 3.4   Alliance Engine

Alliance formation is achieved through the following sequence of steps:
1. The agent willing to participate in an alliance sets the allies parameter to true in its advertisement message which is subsequently sent to advertisement repository.

2. The condition checker, responsible for scanning the advertisement repository for potential buyers and sellers, will also search for potential allies.
3. When allies are found, the condition checker notifies the alliance engine by sending it relevant information such as agent id and product id etc.
4. The alliance engine then informs the agents about their potential allies.
5. The agents then negotiate with one another on the following: (a) Weight of each issue and (b) Sharing of cost for issue
6. The negotiation system results into the formation of new agent responsible for negotiating on the behalf of the parties in the alliance.
7. This new alliance agent now registers with the advertisement repository following the steps described in 3.

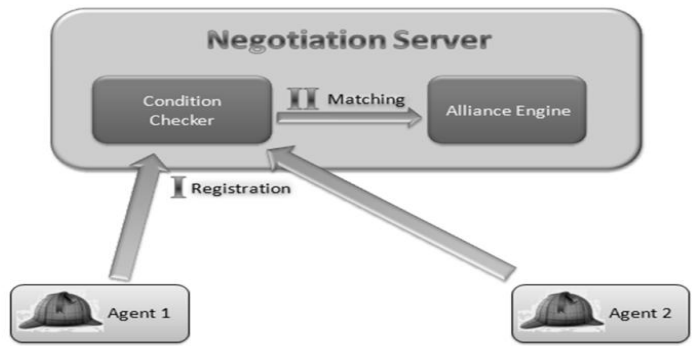

**Figure 2  Alliance Formation (I)**

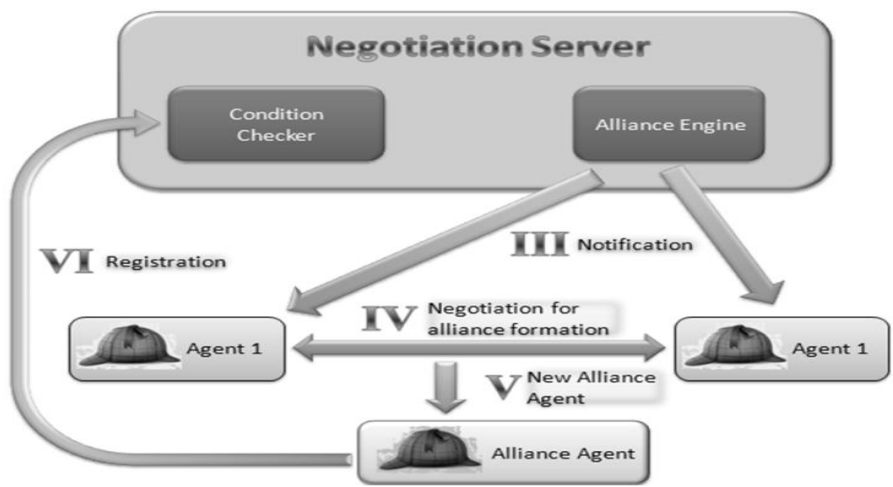

**Figure 3 Alliance Formation (II)**

### 3.5 Agent Architecture

The agent creates a master coordinator that inspects and controls all the coordinators present in that agent for initializing and finalizing the negotiation by generating a coordinator object for each agent with which it is communicating. The coordinator object produces thread for each issue. The coordinator is to be accounted for interacting with the coordinators of other agents by means of threads and continues the negotiation by generation of offers and counter-offers. The coordinator concurrently executes negotiation over multiple issues by using unique threads for each issue. If the offer generated for every single issue is agreed upon the negotiation is successful. Otherwise the threads which have been acknowledged with rejection reconsider their initial offers and regenerate their counter-offers based on their utility functions.

### 3.6 Actual Negotiation Process

The negotiation process begins with the initial offer proposed by both the parties involved in the negotiation i.e. the buyer agent and the seller agent. The seller agent makes his initial offer in which it describes the maximum price of the product. The buyer agent, on the other hand, considers its offer and compares it to its initial minimum price that it has fixed for the negotiation. If the prices match the two parties concur and the deal is accepted otherwise the buyer agent rejects the offer made by the seller agent and instead presents its counter –offer that lists down its minimum price.

The next round of negotiation occurs with the seller agent generating another counter-offer with reduced price in relevance to its initial offer. The buyer agent either accepts the deal or it again generates another counter-offer that has higher prices than its prior offer. Hence both the parties gradually try to converge and come down to an agreement that is amicable. In case of exceeding the time duration assigned for one negotiation settlement the negotiation is aborted.

#### 3.6.1 Evaluation of utility function

For any product the non-functional attributes are those which are judgmental in determining the overall functionality of the product and do not focus on the specific issues. For example, the price (actual cost i.e., the expenses incurred for initial investments for deployment of the product and cost with margin i.e., the final price of a product generated considering the inclusion of profits over the actual cost), the ease of use, updates, time period etc. The functional attributes on the other hand, thoroughly specifies every feature and calculates their minimum and maximum utilities. (1)Minimum utilities: It is the product of all the non-functional attributes and the actual cost. (2) Maximum utilities: It is the product of all the non-functional attributes and the cost with margin. A product with multi-issues has weights i.e., priorities assigned to each issue which signifies their relative importance in the overall valuation of the product. The utility function mathematically depicts the relative usefulness of each attribute (issue) rather than specifying bounded ranges of discrete values to weights of the attributes, the user has flexibility of extending or diminishing the range of values in accordance to the requirement of particular situation. The utility is calculated as follows:

$$U_{min} = (\prod non-functional\ attributes) \times Actual\ Cost$$

$$U_{max} = (\prod non-functional\ attributes) \times Cost\ with\ margin$$

The overall utility of the product is determined by calculating the maximum and minimum payoff as follows:

$$Minimum\ Payoff = \sum U_{min}$$

$$Maximum\ Payoff = \sum U_{max}$$

For the acceptance of an offer, the offered cost of each particular issue should be greater than or equal to the minimum cost.

### 3.6.2 Generation of counter-offers

Assuming that the negotiation process commences with an offer generated by the buyer, we have an initial offer statement with included prices for each attribute. The seller inspects the offer, checking the prices of each issue in succession. It then calculates the utility of each issue with respect to the offered prices and verifies whether prices are acceptable. For unacceptable offers, the following steps are to be followed:

1. Decrease utility as follows:

$$U_{new} = U_{old}(1 - \frac{\lambda' t}{w})$$

where $U_{new}$ = new utility
$U_{old}$ = old utility
$t$ = no. of rounds
$w$ = weight
$\lambda'$ = penalty

2. Derive a new value of $\lambda$ (penalty according to which the utilities can be varied.
3. The coordinator is responsible for calculating the new value of $\lambda$. It takes into consideration all the attributes on which an agreement has not been reached and then depending upon number of rounds left and the weights, it derives a new value of $\lambda$ as shown below:

$$\sum_{x_i\ not\ accepted} x_i(\lambda)$$

$$= \frac{\sum_{U_{i_{max \, not \, accepted}}} U_{i_{max}} - \sum_{U_{i_{min \, not \, accepted}}} U_{i_{min}}}{\text{no. of rounds}}$$

where, $x_{i_{not \, accepted}}$ = cost of attribute i on which agreement has not yet been reached
$w_i$ = weight of attribute i
$\lambda$ = proportionality constant
$U_{i_{max}}$ = maximum utility of attribute i
$U_{i_{min}}$ = minimum utility of attribute i

4. This value of $\lambda$ is sent to the threads on which negotiation is still in progress.
5. It then seals the offers that are acceptable on a temporary settlement, while generating counter-offers for the rejected issues.
6. A time limit counter is updates after every round of negotiation. If the counter exceeds a certain limit, which has been considered as the maximum number of permissible negotiation rounds, the negotiation gets terminated.
7. Till the settlement has not been reached and the time limit counter not exceeded, the counter-offers are generated.

Before negotiation can commence, certain values have to be provided to all the agents to help them carry out the negotiation. The buyer needs values for (1) actual cost, (2) cost with margin, (3) weight of each attribute, as well as (4) some non-functional attribute which will help in calculating the utility associated with the functional attributes. On the other hand, the buyer would only know overall minimum and maximum cost. Value of the non –functional attributes may or may not be supplied to the buyer. In case it is not supplied, these values are taken from the buyer agent making a request to the negotiation system. The values of actual cost and the cost with margin are derived from the maximum and minimum cost by dividing them with number of attributes. The weights may be supplied; however, if they are not available then all the attributes are assigned the same weight. The agent might have the option of selecting different weights for attributes depending on negotiation history.

### 3.6.3 Negotiation Fulfillment Criteria

For the entire approval of a product we need to reach an agreement on every single issue. For sealing the deal, both the agents i.e. the buyer and the seller, exchange finalize () messages. In case an agent is negotiating with more than one agent simultaneously, then the agreement is finalized with the agent who offers the most favorable price (minimum price for buyer and maximum price for seller).

## 4 Salient Features

Our model presents the following salient features:
1. Concurrent and parallel negotiation on each issue using multi-threading.

2. Multi-party and multi-issue negotiation.
3. For the generation of offers and counter-offers, the weight of each issue and utility function are taken into consideration.
4. The non-functional attributes are utilized in the evaluation of varying range of utilities.
5. Enhancements such as hierarchical alignment of the issues in restriction of the negotiation instances to a definite time limit have been implemented.
6. Addition of validity counters to advertisements to kill the advertisements on failure to start negotiation with the specified maximum time interval.
7. Formation of alliances amongst agents with mutual agreements on certain issues.
8. The ability to modify or enhance the internal logic of the agent through the use of plug-ins by importing external components containing the mathematical and operational logic.
9. The concept of fault tolerance has been suggested.

## 5   Future Work

This negotiation model has the scope of getting implemented to form into a full-fledged autonomic negotiation system. Apart from the key database i.e. the advertisement repository, the negotiation history is a secondary database in our model which keeps records of all the messages exchanged among agents for every successful or failed negotiation scenario. The concept is based on the application of Artificial Intelligence in learning by experience wherein the patterns for a successful negotiation are derived and the interface suggests techniques or behavioral patterns in order to maintain the consistency for a successful negotiation by varying the weights associated with each issue. It can also provide pertinent support in identifying the behavioral patterns of the negotiating agents and classifying them into specific groups of agent's viz. conceder, linear, tough etc. The generation of counter offers requires further mathematical treatment. Currently, it is affiliated to a strict linear progression and can have the ability for switching its strategy dynamically and follow other mathematical progressions such as the geometric.

## 6   Conclusion

This paper presents an enhanced negotiation model which performs concurrent and parallel negotiation utilizing the concepts of multithreading thus allowing multi party multi issue negotiation. It provides the opportunity of forming alliances among the agents with converging objectives. External components containing mathematical, logical and business rules can be plugged into the well defined interface of an agent's coordinator thus providing modularity. Utilization of non functional attributes in the evaluation of the relative usefulness of the attribute has been implemented. Hence this paper paves the way towards an improved approach of the negotiation scenario.